\documentclass[twocolumn,times,final,authoryear]{elsarticle}

\usepackage{framed,multirow}

\usepackage{amssymb}
\usepackage{latexsym}
\usepackage{amsmath}

\usepackage{url}

\usepackage[breaklinks,colorlinks]{hyperref}
\usepackage{url}
\usepackage{booktabs}
\usepackage[dvipsnames]{xcolor}
\definecolor{newcolor}{rgb}{.8,.349,.1}
\usepackage{subcaption}

\newcommand{\std}[1]{\scriptsize{$\pm$#1}}

\journal{Pattern Recognition Letters}

\begin{document}

\setcounter{page}{1}

\begin{frontmatter}

\title{Anatomical foundation models for brain MRIs}

\author[unito]{Carlo Alberto {Barbano}\corref{cor1}} 
\cortext[cor1]{Corresponding author.}
\ead{carlo.barbano@unito.it}

\author[unito]{Matteo {Brunello}} %
\ead{matteo.brunello@edu.unito.it}

\author[neurospin]{Benoit {Dufumier}}
\ead{benoit.dufumier@cea.fr}

\author[unito]{Marco {Grangetto}}
\ead{marco.grangetto@unito.it}

\author{for the Alzheimer's Disease Neuroimaging Initiative}\fnref{cor2}
\fntext[cor2]{Data used in preparation of this article were obtained from the Alzheimer's Disease Neuroimaging Initiative (ADNI) database (adni.loni.usc.edu). As such, the investigators within the ADNI contributed to the design and implementation of ADNI and/or provided data but did not participate in the analysis or writing of this report. A complete listing of ADNI investigators can be found at: \url{http://adni.loni.usc.edu/wp-content/uploads/how_to_apply/ADNI_Acknowledgement_List.pdf}}

\affiliation[unito]{
            organization={University of Turin},%
            city={Turin},
            country={Italy}}

\affiliation[neurospin]{
    organization={GAIA, BAOAB, NeuroSpin, CEA Saclay},
    city={Saclay},
    country={France}
}

\begin{abstract}
Deep Learning (DL) in neuroimaging has become increasingly relevant for detecting neurological conditions and neurodegenerative disorders. One of the most predominant biomarkers in neuroimaging is represented by brain age, which has been shown to be a good indicator for different conditions, such as Alzheimer's Disease. Using brain age for weakly supervised pre-training of DL models in transfer learning settings has also recently shown promising results, especially when dealing with data scarcity of different conditions. On the other hand, anatomical information of brain MRIs (e.g. cortical thickness) can provide important information for learning good representations that can be transferred to many downstream tasks. In this work, we propose AnatCL, an anatomical foundation model for structural brain MRIs that i.) leverages anatomical information in a weakly contrastive learning approach, and ii.) achieves state-of-the-art performances across many different downstream tasks. To validate our approach we consider 12 different downstream tasks for the diagnosis of different conditions such as Alzheimer's Disease, autism spectrum disorder, and schizophrenia. Furthermore, we also target the prediction of 10 different clinical assessment scores using structural MRI data. Our findings show that incorporating anatomical information during pre-training leads to more robust and generalizable representations. Pre-trained models can be found at: \url{https://github.com/EIDOSLAB/AnatCL}.

\end{abstract}

\begin{keyword}
Brain foundation models \sep Contrastive Learning \sep Neuroimaging
\end{keyword}

\end{frontmatter}

\section{Introduction}

Magnetic Resonance Imaging (MRI) is the foundation of neuroimaging, providing detailed views of the brain structure and function that are crucial for diagnosing and understanding various neurological conditions. However, the complexity and high dimensionality of MRI data present significant challenges for automated analysis. Traditional machine learning methods often require extensive manual annotation, which is both time-consuming and expensive. Recently, contrastive deep learning techniques have emerged as powerful tools for extracting meaningful information from complex data with minimal requirement for annotations, showing promise in improving the automated analysis of neuroimaging datasets.
\begin{figure}
    \centering
    \includegraphics[width=0.8\linewidth,trim=10 20 10 10,clip]{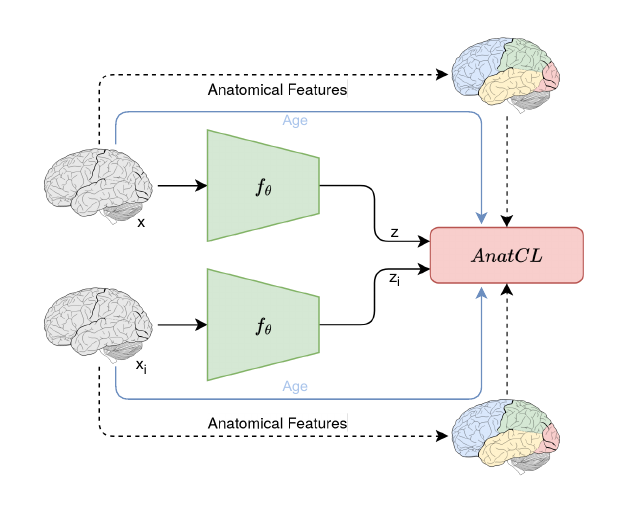}
    \caption{\textbf{We propose AnatCL}, a contrastive learning approach that leverages patient's metadata (i.e., age) alongside anatomical information derived from structural MRIs, in order to learn meaningful and general representations. Given two input samples $x$ and $x_i$, the objective of AnatCL is to learn embeddings $z$ and $z_i$, such that their distance in the representation space is proportional to the difference in age and brain anatomy between the two samples.}
    \label{fig:anatcl-summary}
\end{figure}
Contrastive learning methods, particularly those leveraging weakly supervised loss functions, have been shown to be effective in creating models that learn robust representation spaces~\citep{dufumier2021contrastive, dufumier2024exploring}. These methods typically utilize metadata, such as the patient's age, to guide the learning process, to align similar data points (e.g., similar age) and distinguish dissimilar ones in the learned representation space. While this approach has yielded promising results, it may fall short of fully capturing the multifaceted information inherent in MRI data and the anatomical complexity of the brain, as it primarily relies on age alone.

In this paper, we extend the capabilities of weakly supervised contrastive learning approaches by incorporating additional anatomical measures into the loss formulation.
We propose AnatCL, a contrastive learning framework that improves upon previous works leveraging patient age as proxy metadata~\citep{dufumier2021contrastive,barbano2023contrastive}, to also include information derived from anatomical measures (such as cortical thickness), computed on region of interests (ROIs) of common atlas such as Desikan-Killiany~\citep{desikan2006automated} or Destrieux~\citep{destrieux2010automatic}. Specifically, inspired by the MIND anatomical similarity metric~\citep{sebenius2023robust}, we propose a modified contrastive learning loss function that integrates patient age with a targeted selection of three anatomical features: Mean Cortical Thickness, Gray Matter Volume, Surface Area. This selective integration was designed to assess the effectiveness of using potentially more informative anatomical features in conjunction with demographic data.

We propose and test two versions of our framework: a local version and a global version. %
The local version aims to capture the localized anatomical variations between specific ROIs (i.e., computing a cross-region similarity across all measures), while the global version aggregates these features across the entire brain to provide a more holistic view of the general anatomy (i.e., computing a cross-measure similarity across all regions). A graphical overview of AnatCL is provided in Fig.~\ref{fig:anatcl-summary}.
We pre-train our models using the proposed AnatCL loss on the OpenBHB dataset, a large collection of healthy samples, and then we test the resulting models on different downstream tasks across a variety of publicly available datasets (Sec.~\ref{sec:experimental-data}). Our results show that in several downstream tasks, AnatCL achieves better results than traditional pre-training approaches based on brain age prediction only. These findings suggest that, in contrastive learning frameworks, incorporating anatomical features alongside the patient's age can enhance the expressiveness of the learned representations, leading to the development of models that learn more meaningful and generalizable representation spaces. Additionally, we also perform experiments to assess whether the embeddings learned using anatomical information can be leveraged for predicting clinical assessment scores, cognition, and psychopathology of the subjects (e.g., IQ, depression scores, and abnormal involuntary movements). To the best of our knowledge, this is one of the first works attempting to tackle this objective, which is typically performed on functional MRI data~\citep{kong2023comparison}. In summary:

\begin{enumerate}
    \item We propose a novel weakly contrastive learning formulation that takes into account multiple features and metadata for guiding the learning process;
    
    \item We propose AnatCL, a foundation model based on both anatomical and age features to build robust and generalizable representations of structural neuroimaging data;
    
    \item We perform extensive validation across 12 different deep phenotyping tasks and the prediction of 10 different clinical assessment scores.
\end{enumerate}

\section{Related Works}

Many works have explored the potential of machine learning and deep learning in neuroimaging. Here, we review the most popular approaches, distinguishing relevant topics for our work. First, we start with standard machine learning approaches, then we explore self-supervised methods, and finally, we present state-of-the-art contrastive approaches for neuroimaging.

\paragraph{Machine learning on structural brain imaging} Current literature has been mainly devoted to segmentation tasks for brain anatomical data, such as brain tumor segmentation~\citep{biratu2021survey} or tissue segmentation~\citep{dora2017state}, in a supervised setting requiring annotated maps. The deep neural network U-Net~\citep{ronneberger2015u} and its extension nnU-Net~\citep{isensee2021nnu} were very successful at such tasks. Another body of literature tackled phenotype and clinical prediction tasks from structural imaging, such as age prediction~\citep{franke2019ten}, Alzheimer’s disease detection~\citep{wen2020convolutional}, cognitive impairment detection, or psychiatric conditions prediction~\citep{nunes2020using, chand2020two}. Current approaches mostly involve traditional machine learning models such as SVM or penalized linear models trained in a supervised fashion to predict the targeted phenotypes. Since most studies only include a few hundred subjects in their datasets, larger models introducing more parameters (such as neural networks) do not necessarily translate into better performance~\citep{dufumier2024exploring, schulz2020different, he2020deep}, probably because of over-fitting.

\paragraph{Self-supervised learning on brain imaging} Self-supervised learning has been popularized by its impressive performance in Natural Language Processing (NLP) using auto-regressive models trained on web-scale text corpus and mainly deployed on generation tasks~\citep{devlin2018bert, brown2020language}. Since it does not require a supervised signal for training, it is particularly appealing when large-scale data are available, and the annotation cost is prohibitive. Its applicability on neuroimage data is still unclear and very few studies have tackled this problem so far~\citep{huang2023self}. \cite{dufumier2021contrastive} demonstrated that contrastive learning integrating subjects meta-data can improve the classification of patients with psychiatric conditions. \cite{thomas2022self} demonstrated that learning frameworks from NLP, such as sequence-to-sequence autoencoding, causal language modeling, and masked language modeling, can improve mental state decoding from brain activity recorded with functional MRI. 

\paragraph{Contrastive Learning on brain imaging} Recently, interest in contrastive learning approaches for brain age prediction has risen. It has been demonstrated that contrastive learning performs better than traditional methods for neuroimaging~\citep{dufumier2024exploring}. Among different works, we can identify y-Aware~\citep{dufumier2021contrastive} and ExpW~\citep{barbano2023contrastive}. These works propose to utilize contrastive learning for continuous regression tasks by employing a kernel to measure similarity between different samples. In these works, age prediction is used as pre-training task, followed by transfer learning on specific diseases. However, they are limited to only considering age as pre-training feature. In this work, we extend this formulation to work with multiple attributes.

\section{Method}

In this section, we present the proposed method AnatCL. To introduce the adequate tools and concepts for understanding our work, we first provide an introduction to contrastive learning and weakly contrastive learning. Then, we present AnatCL.

\subsection{Background}

Contrastive learning (CL), either self-supervised as in SimCLR~\citep{chen2020simple} or supervised as in SupCon~\citep{khosla2020supervised}, leverages discrete labels (i.e., categories) to define positive and negative samples. Starting from a sample $x$, called the \emph{anchor}, and its latent representation $z = f(x)$, CL approaches aim at aligning the representations of all positive samples to $z$, while repelling the representations of the negative ones.
CL is thus not adapted for regression (i.e., continuous labels), as a hard boundary between positive and negative samples cannot be determined. 
Recently, there have been developments in the field of CL for tackling regression problems, especially aimed at brain age prediction, such as y-Aware~\citep{dufumier2021contrastive} and ExpW~\citep{barbano2023contrastive}, from brain MRIs. These approaches can be broadly categorized as weakly contrastive, as brain age is used as pre-train task. 

\begin{figure}
    \centering
    \includegraphics[width=0.8\linewidth]{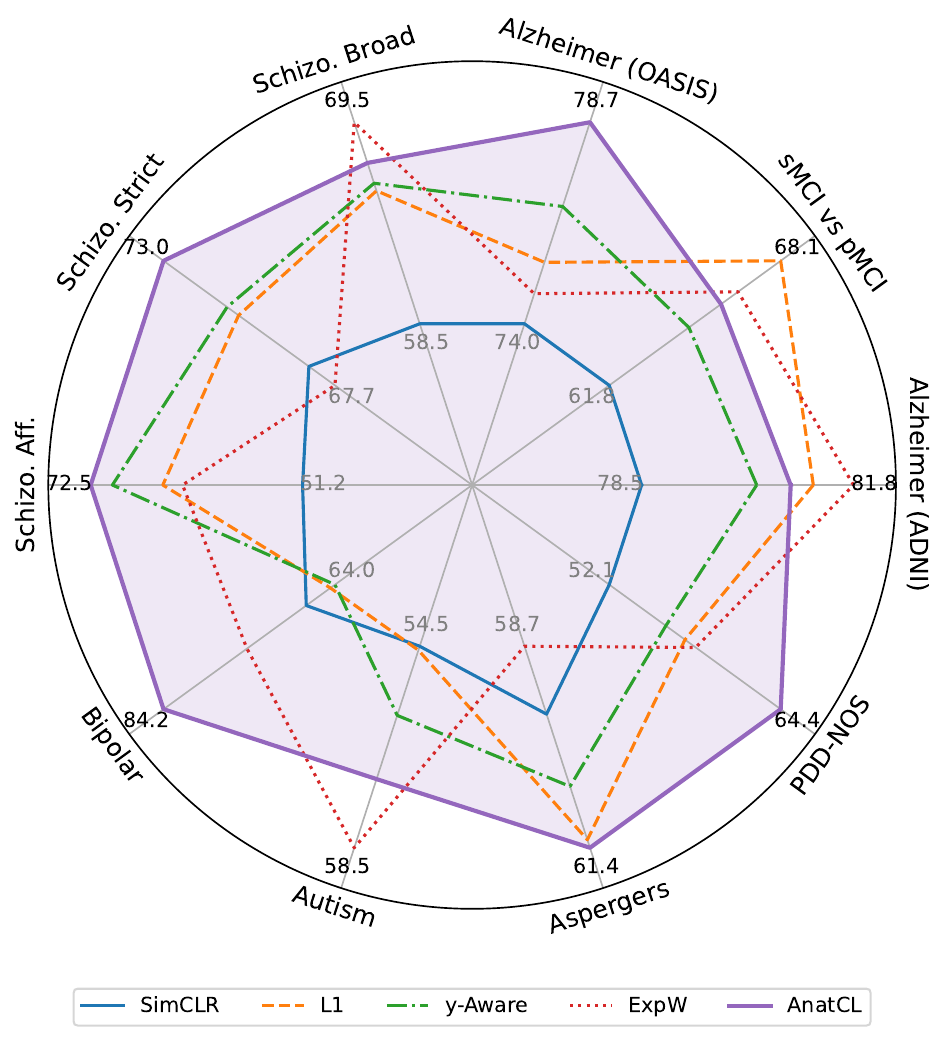}
    \caption{\textbf{Graphical evaluation of AnatCL (local and global) on all deep phenotyping tasks considered.} Overall, AnatCL achieves higher results more consistently than any other baseline, with notable performance on ASD and Schizophrenia.}
    \label{fig:results-polar}
\end{figure}

\subsection{Weakly contrastive learning with age}

To tackle the issue of continuous labels, weakly contrastive approaches employ the notion of a degree of ``positiveness'' between samples~\citep{barbano2023contrastive}. This degree is usually defined by a kernel function $w_i = \mathcal{K}(y - y_i)$, where $0 \leq w_i \leq 1$ is computed by a Gaussian or a Radial Basis Function (RBF) kernel and $y, y_i$ are the continuous attributes of interest (e.g. age). 
The goal of weakly contrastive learning is thus to learn a parametric function $f: \mathcal{X} \rightarrow \mathbb{S}^d$ that maps samples with a high degree of positiveness ($w_i \sim 1$) close in the latent space and samples with a low degree ($w_i \sim 0$) far away from each other. 
y-Aware~\citep{dufumier2021contrastive} proposes to do so by aligning positive samples  with a strength proportional to the degree, i.e.:

\begin{equation}
     \mathcal{L}^{y-aware}= - \sum_i \frac{w_i}{\sum_j w_j} \log \left( \frac{\exp(sim(z, z_i))}{\sum_{t=1}^N \exp(sim(z, z_t))}  \right)
     \label{eq:yaware}
\end{equation}

where $z = f(x)$, $z_i = f(x_i)$, $z_t = f(x_t)$, and $sim$ is a similarity function (e.g. cosine). A similar approach is also adopted by ExpW~\citep{barbano2023contrastive}. A limitation of these approaches is the reliance on one single attribute (age) to determine the alignment strength. They are thus not suited to leverage multiple attributes (e.g., anatomical measurements of the brain). Our proposed approach aims to solve this issue by extending the weakly contrastive formulation to include multiple attributes.

\subsection{Anatomical Contrastive Learning}
\label{sec:anatcl}

We propose a novel formulation for weakly contrastive learning to employ anatomical attributes derived from the MRI scan, which we call AnatCL. The features we consider are available in the OpenBHB dataset (see Sec.~\ref{sec:experimental-data}) for both Desikan-Killany and Destrieux atlases, composed of 68 and 148 ROIs respectively.
Inspired by MIND~\citep{sebenius2023robust}, we consider cortical thickness (CT), gray matter volume (GMV), and surface area (SA). Additional measures are available in the dataset; however, we did not achieve better results when using them (see supp. material).
To include these measurements inside AnatCL, we propose two different formulations, global and local, based on how the similarity across two brain scans is computed.

\paragraph{Local Descriptor}
Given a sample $x$, we consider the set of local descriptors $\Psi(x) = \{\psi^1(x),\dots\,\psi^K(x)\}$, where $\psi^k(x) \in \mathbb{R}^N$ for $k \in \{1,\dots,K\}$, $K$ is the number of region of interest (ROIs) in the chosen atlas (68 for Desikan, 148 for Destrieux), and $N$ is the number of features (i.e., 3).
Each local descriptor is thus a vector composed of the three anatomical measurements CT, GMV, and SA, corresponding to a specific ROI of the atlas.
To compute a degree of positiveness $\alpha_i$ based on the local descriptors of two samples $x$ and $x_i$, we consider the average cross-region similarity as:
\begin{equation}
    \alpha_i = \frac{1}{K}\sum_{k=1}^{K} s(\gamma(\psi^k), \gamma(\psi_i^k))
\end{equation}
\noindent where, for brevity $\psi^k = \psi^k(x)$ and $\psi_i^k = \psi^k(x_i)$, $s$ can either be a similarity function or a kernel as in~\cite{dufumier2021contrastive}, and $\gamma(\psi)$ is a normalization function that normalizes each component of $\psi$ to a standard range $[0,1]$. In this work, for simplicity, we choose to employ cosine similarity as $s$. By plugging $a_i$ into Eq.~\ref{eq:yaware}, we obtain the local variant of AnatCL (omitted the normalization terms $1/\sum_j \alpha_k$ and $1/K$ for brevity):
\begin{equation}
\begin{split}
    \mathcal{L}_{AnatCL-L3} = \qquad\qquad\qquad\qquad\qquad\qquad\qquad\qquad \\
    = -\sum_i\sum_{k=1}^{K} s(\gamma(\psi^k), \gamma(\psi_i^k))\log \left( \frac{\exp(sim(z, z_i))}{\sum_{t}^N \exp(sim(z, z_t))}  \right)
    \label{eq:anatcl-local}
\end{split}
\end{equation}

\paragraph{Global Descriptor}
Taking a complementary approach, we now consider global anatomy descriptors $\omega^j(x) \in \mathbb{R}^{K}$ for the entire brain with $j \in \{CT,GMV,SA\}$, that contain the values across all $K$ regions for each anatomical measurement. Similarly to above, in order to compute a degree of positiveness $\beta_i$ between two samples $x$ and $x_i$ given their global anatomical descriptors $\Omega(x) = \{\omega^{CT}(x),\omega^{GMV}(x),\omega^{SA}(x)\}$ and $\Omega(x_i) = \{\omega^{CT}(x_i),\omega^{GMV}(x_i),\omega^{SA}(x_i)\}$ we compute the cross-measurement similarity (global descriptor) as:
\begin{equation}
    \beta_i = \frac{1}{3}\sum_{j} s(\omega^j, \omega_i^j) \quad j \in \{CT, GMV, SA\}
\end{equation}
\noindent where, similarly to before, $\omega^j = \omega^j(x)$ and $\omega_i^j = \omega^j(x_i)$. This time we do not need to normalize $\omega^j$ and $\omega_i^j$ as $s$ is evaluated between features of the same type, so the scale of the values is comparable. Also in this case, we employ a cosine similarity function for $s$. As before, by plugging $\beta_i$ in Eq.~\ref{eq:yaware} we obtain the AnatCL global loss (omitting the normalization terms):
\begin{equation}
\begin{split}
    \mathcal{L}_{AnatCL-G3} = \qquad\qquad\qquad\qquad\qquad\qquad\qquad\qquad \\
    = -\sum_i\sum_{j} s(\omega^j, \omega_i^j)\log \left( \frac{\exp(sim(z, z_i))}{\sum_{t}^N \exp(sim(z, z_t))}  \right)
    \label{eq:anatcl-global}
\end{split}
\end{equation}

\subsubsection{Final objective function}

The final objective function is a combination of our proposed AnatCL loss (either local or global) and the y-Aware loss which considers age. Our final formulation is thus:

\begin{equation}
    \mathcal{L} = \lambda_1\mathcal{L}_{AnatCL} + \lambda_2\mathcal{L}_{age}
\end{equation}

\noindent where $\mathcal{L}_{age}$ is defined as in Eq.~\ref{eq:yaware} and $\mathcal{L}_{AnatCL}$ is defined by either Eq.~\ref{eq:anatcl-local} or Eq.~\ref{eq:anatcl-global}. The weights $\lambda_1$ and $\lambda_2$ control the contribution of each loss term independently. 
It is worth noting that the considered anatomical features in AnatCL can be computed directly from the MRI with standard tools such as FreeSurfer\footnote{\url{https://surfer.nmr.mgh.harvard.edu/}}, thus our method does not require any additional label in the dataset.

\section{Experiments}
In this section, we present the experimental data that we employ, and the experiments that we perform to assess the performance of AnatCL. We compare AnatCL to different baselines, composed of state-of-the-art approaches based on contrastive learning in several downstream tasks. An overview of downstream performance of AnatCL is provided in Fig.~\ref{fig:results-polar}. From the plot, we can notice that AnatCL obtains state-of-the-art results in more tasks than any of the other methods.

\subsection{Experimental data}
\label{sec:experimental-data}

We use a collection of T1-weighted MRI scans comprising 7,908 different individuals, for a total of 21,155 images. The experimental data is gathered from different publicly available datasets, and we target 5 different neurological conditions: healthy samples (OpenBHB~\citep{dufumier2022openbhb}), Alzheimer's Disease (ADNI~\citep{petersen2010alzheimer} and OASIS-3~\citep{lamontagne2019oasis}), schizophrenia (SchizConnect~\citep{wang2016schizconnect}) and Autism Spectrum Disorder (ABIDE I~\citep{ABIDE}).
Tab.~\ref{tab:downstreams-summary} provides an overview of all the downstream tasks that we explore in this work. 

\paragraph{OpenBHB} OpenBHB is a recently released dataset that aggregates healthy control (HC) samples from many public cohorts (ABIDE 1, ABIDE 2, CoRR, GSP, IXI, Localizer, MPI-Leipzig, NAR, NPC, RBP). Every scan comes from a different subject, for a total of 3984 MRI images. We use OpenBHB as the pre-training dataset for our method.
Besides the structural scans and patients information, OpenBHB also provides 7 anatomical measures: cortical thickness mean and std, gray matter volume, surface area, integrated mean, gaussian curvature index, and intrinsic curvature index~\citep{dufumier2022openbhb}.

\paragraph{ADNI and OASIS-3} The Alzheimer's Disease Neuroimaging Initiative (ADNI)\footnote{Data used in the preparation of this article were obtained from the Alzheimer's Disease Neuroimaging Initiative (ADNI) database (\url{adni.loni.usc.edu}).} is one of the largest and most popular datasets of MRIs for Alzheimer's Disease (AD). In our study, we include all phases ADNI-1, ADNI-2, ADNI-GO and ADNI-3 amounting to 633 HC, 712 mild cognitive impairment (MCI) and 409 AD patients, for a total of 14,385 MRI images. We downloaded all raw acquisition in the DICOM format and performed VBM preprocessing with CAT12~\citep{gaser2022cat}. We also collected all clinical information and processed it in order to distinguish between stable and progressive MCI. Additionally, we also employ the OASIS3 dataset, including 685 patients (88 AD cases and 597 control cases), for a total of 1354 images.

\paragraph{SchizConnect} For schizophrenia, we employ SchizConnect, a large dataset of MRIs. We include structural MRIs for a total of 383 patients, divided into 180 HC, 102 with schizophrenia (broad), 74 with schizophrenia (strict), 11 schizoaffective patients, and 9 with bipolar disorder, for a total of 383 MRIs.

\paragraph{ABIDE I}  The Autism Brain Imaging Data Exchange (ABIDE) dataset is one of the most popular datasets of patients affected by conditions of Autism Spectrum Disorder (ASD). We leverage data from ABIDE-I, including structural MRIs for a total of 1102 patients, divided into 556 HC, 339 patients with autism, 93 patients with Asperger's Syndrome, and 7 with pervasive developmental disorder not otherwise specified (PDD-NOS). The total number of images is 1102.

\paragraph{Clinical assessment scores and phenotypes} In addition to predicting the diagnosis, we also consider some relevant clinical assessments included in the data, as reported in Tab.~\ref{tab:downstreams-summary}. For SchizConnect, we consider the Abnormal Involuntary Movement Scale (AIMS) evaluated in three assessments (overall, upper body, lower body), a Depression score based on the Calgary Scale, Handedness information, and GAIT measurements with the Simpson-Angus-Scale (SAS). For ABIDE, we also consider handedness and three IQ scores, namely Full Scale IQ (FIQ), Visual IQ (VIQ), and Performance IQ (FIQ) measured with the Wechsler Abbreviated Scale (WASI). To the best of our
knowledge, this is the first work attempting to tackle
this objective.

\begin{table}
    \centering
    \caption{\textbf{Summary of downstream tasks} (12) and clinical assessment scores (10) considered in the study.}
    \resizebox{\linewidth}{!}{%
    \begin{tabular}{l l}
    \toprule
    \textbf{Dataset} & \textbf{Task / Condition} \\
    \midrule
    \textbf{OpenBHB} & Age (HC), Sex   \\
            & \\
    \textbf{ADNI}    & Alzheimer's Diseas, \\
            & sMCI vs pMCI \\
            & \\
    \textbf{OASIS-3} & Alzheimer's Disease \\
            & \\
    \textbf{SchizConnect} & Schizo. Broad \& Strict, \\
                 & Bipolar Disorder, \\
                 & Schizoaffective \\
                 & \\
    \textbf{ABIDE I} & Autism, Aspergers, \\
            & PDD-NOS \\
    \bottomrule
    \end{tabular}
    \begin{tabular}{l l}
    \toprule
    \textbf{Dataset} & \textbf{Phenotype}  \\
    \midrule
    \textbf{SchizConnect}  & AIMS Overall Severity  \\
                  & AIMS Upper Body \\
                  & AIMS Lower Body \\
                  & Depression \\
                  & Handedness \\
                  & SAS GAIT \\
                  & \\
    \textbf{ABIDE 1} & Handedness \\
            & FIQ (WASI)\\
            & VIQ (WASI) \\
            & PIQ (WASI) \\
    & \\
    & \\
    \bottomrule
    \end{tabular}
    }
    \label{tab:downstreams-summary}
\end{table}

\paragraph{Data preparation and preprocessing}
All the images underwent the same standard VBM preprocessing, using CAT12~\citep{gaser2022cat}, which includes non-linear registration to the MNI template and gray matter (GM) extraction. The final spatial resolution is 1.5mm isotropic and the images are of size 121 x 145 x 121. The preprocessing is performed using the brainprep package\footnote{\url{https://brainprep.readthedocs.io/en/latest/}}. After preprocessing, we consider modulated GM images, as in~\cite{dufumier2022openbhb}.

\paragraph{Computation of anatomical features}

Anatomical measurements on OpenBHB are released as part of the original dataset. %
For deriving the measurements, the brainprep module was employed, with FreeSurfer version FSL-6.0.5.1, and the Desikan-Killiany and Destrieux atlases as provided in FreeSurfer. We chose to employ CT, GMV, and SA as they are also used in~\cite{sebenius2023robust}; for additional results on other measures, please refer to the supplementary materials.

\subsection{Experimental setup}
\label{sec:experimental_setup}
Our experiments involved pre-training a model using the proposed AnatCL loss on the OpenBHB dataset, followed by testing on various downstream tasks across different datasets.
The experimental settings for the two loss formulations, local (AnatCL-G3) and global (AnatCL-L3), are identical. 
We pre-train two ResNet-18 3D models using VBM-preprocessed images and their corresponding anatomical measures with the proposed formulations.
The training process employs the Adam optimizer with a learning rate of 0.0001 and a decay rate of 0.9 applied after every 10 epochs. The models are trained with a batch size of 32 for a total of 300 epochs. As values of $\lambda_1$ and $\lambda_2$, for simplicity, we use 1.
As standard practice in CL approaches~\citep{chen2020simple,khosla2020supervised}, the contrastive loss is computed on a fully-connected projection head following the encoder, composed of two layers.
To ensure robust evaluation, we perform cross-validation using 5 folds. The results are computed in terms of mean and standard deviation across the 5 folds. %
After the pre-training step we evaluate the models by testing their performance with a transfer learning approach: we extract the latent representations generated by the model using only the encoder of the model (i.e., discarding the fully connected head). For each downstream task, we train different linear classifiers on the extracted representations to assess the model's ability to learn meaningful and generalizable features. 
To run our experiments, we use the Jean Zay and Leonardo clusters (providing NVIDIA V100 and A100 GPUs), with a single training taking around 10h.

\paragraph{Baselines for comparison}
In this work we focus on contrastive learning has it has been demonstrated to work better than non-contrastive method in recent literature~\citep{dufumier2024exploring}. For this reason, we pick recent state-of-the-art contrastive methods in neuroimaging, namely y-Aware~\citep{dufumier2021contrastive} and ExpW~\citep{barbano2023contrastive}. In addition to these recent techniques, we also consider the popular SimCLR framework~\citep{chen2020simple} as a self-supervised contrastive baseline~\citep{tak2024foundation}. For comparison with traditional non-contrastive supervised techniques, we also compare with baseline models trained with L1 loss for brain age prediction~\citep{jonsson2019brain}.

\subsection{Results}

We evaluate the results of AnatCL pre-trained on OpenBHB and tested on different datasets, for several deep phenotyping and clinical assessment scores prediction tasks.

\paragraph{Brain age and sex prediction}
The preliminary downstream tasks we evaluate are brain age prediction and sex classification, on OpenBHB. The results are reported in Tab.~\ref{tab:openbhb-results}.
AnatCL achieves state-of-the-art results on brain age prediction, reaching the lowest absolute error of 2.55 years. It is worth noting that we do not employ any bias-correction method~\citep{delange2020commentary}, but we consider the raw predictions of the model. For sex classification, AnatCL is competitive with ExpW~\citep{barbano2023contrastive}, and improves over the other baselines.

\begin{table}
    \centering
    \caption{\textbf{Results on OpenBHB} in terms of mean absolute error (MAE) on age prediction, and balanced accuracy on sex classification. AnatCL-L3 refers to the local version, while AnatCL-G3 refers to the global version; ``3'' refers to the three measures employed (CT, GVM, SA).}
    \begin{tabular}{l c c}
    \toprule
    \textbf{Method} & \textbf{Age MAE} & \textbf{Sex}  \\
    \midrule
    SimCLR~\cite{chen2020simple} & 5.58\std{0.53} & 76.7\std{1.67} \\
    L1 (age supervised) & 2.73\std{0.14} & 76.7\std{0.67} \\
    y-Aware~\cite{dufumier2021contrastive} & 2.66\std{0.06} & 79.6\std{1.13} \\
    ExpW~\cite{barbano2023contrastive} & 2.70\std{0.06} & \textbf{80.3}\std{1.7} \\
    \midrule
    AnatCL-L3 (Desikan) & \textbf{2.55}\std{0.07} & \underline{80.0}\std{1.0}\\
    AnatCL-L3 (Destrieux) & 2.58\std{0.08} & 80.0\std{1.0} \\
    AnatCL-G3 (Desikan) & 2.56\std{0.06} & 79.0\std{1.0} \\
    AnatCL-G3 (Destrieux) & 2.59\std{0.05} & 80.0\std{1.0}\\
    \bottomrule
    \end{tabular}
    \label{tab:openbhb-results}
\end{table}

\begin{table*}[t]
    \centering
    \caption{\textbf{Deep phenotyping results} across psychiatric (SCZ, ASD)
             and neuro-degenerative (AD) benchmarks in terms of balanced accuracy.}
    \resizebox{\linewidth}{!}{%
    \begin{tabular}{l
                    c c c c | c c c | c c | c}
    \toprule
    & \multicolumn{4}{c|}{\textbf{Schizconnect}}
    & \multicolumn{3}{c|}{\textbf{ABIDE-I}}
    & \multicolumn{2}{c|}{\textbf{ADNI}}
    & \textbf{OASIS-3}\\[-0.2em]
    \textbf{Method}
        & \textbf{SCZ (Broad)}
        & \textbf{SCZ (Strict)}
        & \textbf{Schizoaff.}
        & \textbf{Bipolar}
        & \textbf{Autism}
        & \textbf{Aspergers}
        & \textbf{PDD-NOS}
        & \textbf{HC vs AD}
        & \textbf{sMCI vs pMCI}
        & \textbf{HC vs AD}\\
    \midrule
    SimCLR
        & 58.53\std{3.52} & 68.47\std{8.47} & 51.23\std{11.94} & 67.34\std{9.96}
        & 54.45\std{2.99} & 59.61\std{2.72} & 52.12\std{6.62}
        & 78.47\std{2.51} & 61.77\std{3.85} & 73.97\std{4.98}\\
    L1 (age-sup.)
        & 65.79\std{4.74} & 70.68\std{5.10} & 65.24\std{15.21} & 64.49\std{22.08}
        & 54.53\std{1.79} & \underline{61.33}\std{8.77} & 57.54\std{5.93}
        & \underline{81.20}\std{2.30} & \textbf{68.12}\std{5.42} & 75.40\std{5.40}\\
    y-Aware
        & 66.20\std{4.50} & 71.04\std{2.31} & \underline{70.29}\std{14.73} & 63.95\std{19.81}
        & 55.84\std{3.37} & 60.60\std{9.22} & 56.17\std{10.17}
        & 80.30\std{1.80} & 64.72\std{4.43} & 76.70\std{3.30}\\
    ExpW
        & \textbf{69.53}\std{4.43} & 67.65\std{8.27} & 63.26\std{18.06} & \underline{74.34}\std{18.95}
        & \textbf{58.50}\std{2.41} & 58.68\std{3.82} & \underline{58.31}\std{4.33}
        & \textbf{81.84}\std{2.95} & \underline{66.54}\std{5.64} & 74.67\std{2.87}\\
    \midrule
    AnatCL-L3 (Desikan)
        & 66.03\std{4.06} & 69.86\std{11.20} & 52.72\std{6.64} & 63.12\std{21.50}
        & 53.99\std{2.60} & \textbf{61.44}\std{4.33} & 56.67\std{7.43}
        & 79.71\std{2.51} & 64.78\std{3.57} & 74.26\std{3.32}\\
    AnatCL-L3 (Destrieux)
        & \underline{67.31}\std{5.75} & 70.03\std{7.16} & \textbf{72.47}\std{12.27} & 55.88\std{13.23}
        & 53.14\std{4.20} & 60.99\std{4.94} & 56.63\std{6.74}
        & 79.38\std{3.56} & 64.97\std{3.40} & 77.00\std{2.28}\\
    AnatCL-G3 (Desikan) & 64.31\std{4.43} & \underline{72.46}\std{4.41} & 62.42\std{14.95} & 72.42\std{15.77} & \underline{57.12}\std{2.79} & 58.25\std{4.46} & \textbf{64.39}\std{9.26} & 80.84\std{2.38} & 64.88\std{4.37} & \underline{78.32}\std{4.48}\\
    AnatCL-G3 (Destrieux) & 64.12\std{3.18} & \textbf{73.04}\std{4.42} & 59.76\std{15.20} & \textbf{84.19}\std{7.10} & 55.41\std{4.55} & 58.96\std{7.52} & 58.10\std{4.63} & 80.75\std{3.35} & 65.91\std{2.23} & \textbf{78.66}\std{2.89}\\
    \bottomrule
    \end{tabular}}
    \label{tab:combined-results}
\end{table*}

\paragraph{Schizophrenia and ASD}
We evaluate phenotyping performance on SchizConnect for detecting schizophrenia (broad and strict), schizoaffective, and bipolar patients. Results are reported in Tab.~\ref{tab:combined-results}. With AnatCL, we achieve state-of-the-art performance on three out of four tasks, suggesting that including anatomical information during training can prove useful for these psychiatric conditions.
We also assess detection performance for Autism Spectrum Disorder (ASD) patients, across three categories (autism, aspergers, and PDD-NOS). While AnatCL does not beat the other methods on general autism (but is competitive), we achieve the best results on Aspergers and significantly improve the accuracy for PDD-NOS patients, which is a rarer diagnosis in the data. Overall, AnatCL achieves better results than other baselines most of the time, and is competitive with SOTA results in the remaining instances.

\paragraph{Alzheimer's desease and cognitive impariments}

In Tab.~\ref{tab:combined-results} we also report results for Alzheimer's Disease (AD) detection on both ADNI and OASIS-3. On ADNI, we also assess the discriminative power between stable MCI (sMCI) and progressive MCI (pMCI) patients. While we do not reach the best results overall on ADNI, AnatCL can consistently improve over the self-supervised baseline SimCLR and over the weakly supervised method y-Aware, and reaches overall competitive results. On OASIS-3, AnatCL also achieves the best results in AD detection, in both local and global variants. Overall, AnatCL provides more consistent performance than other methods.

\paragraph{Clinical assessments scores}

\begin{figure*}
    \centering
    \begin{subfigure}[t]{0.48\linewidth}
        \includegraphics[width=\linewidth]{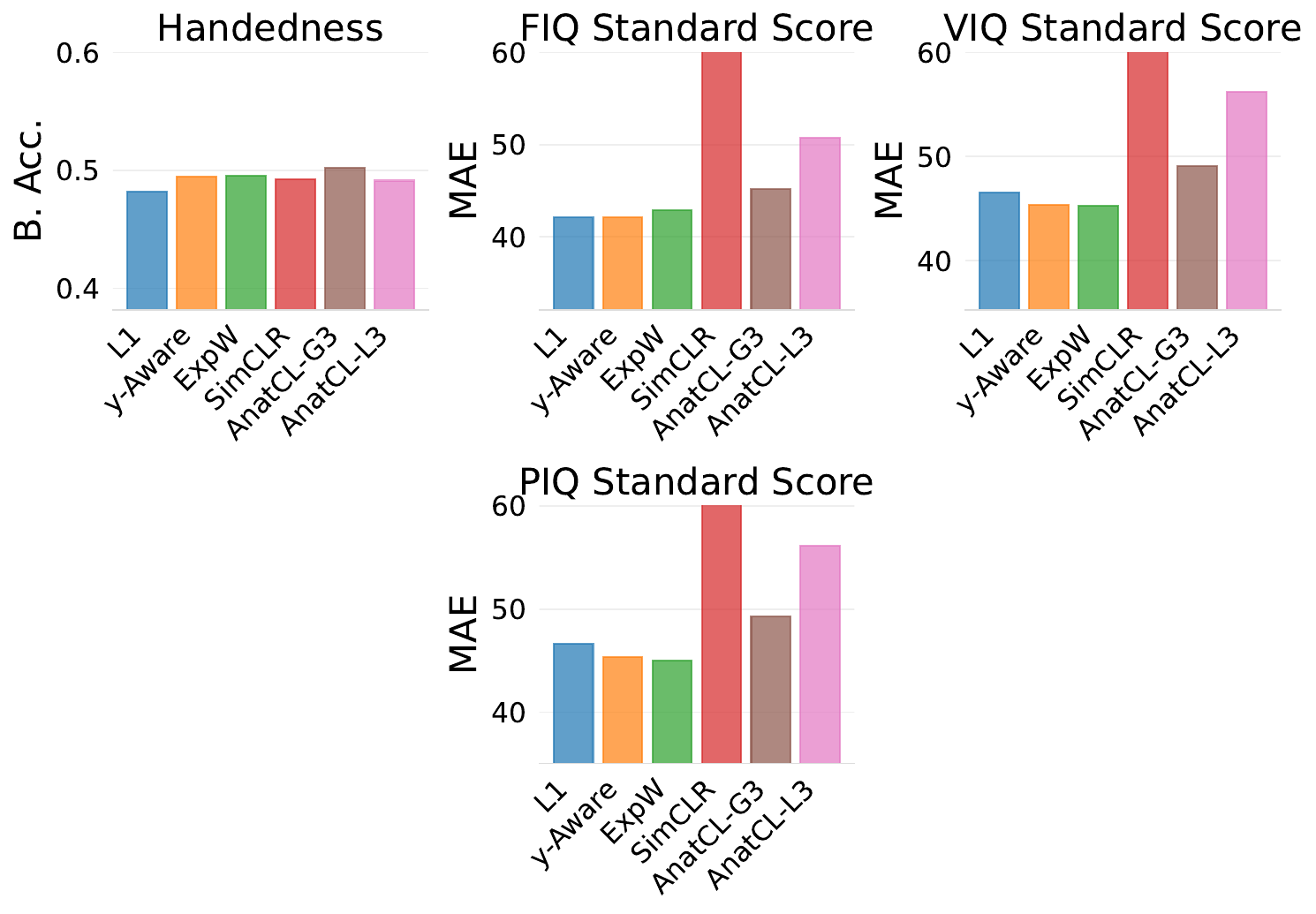}
        \caption{ABIDE-I}
        \label{fig:asd-pheno}
    \end{subfigure}
    \hfill
    \begin{subfigure}[t]{0.48\linewidth}
        \includegraphics[width=\linewidth]{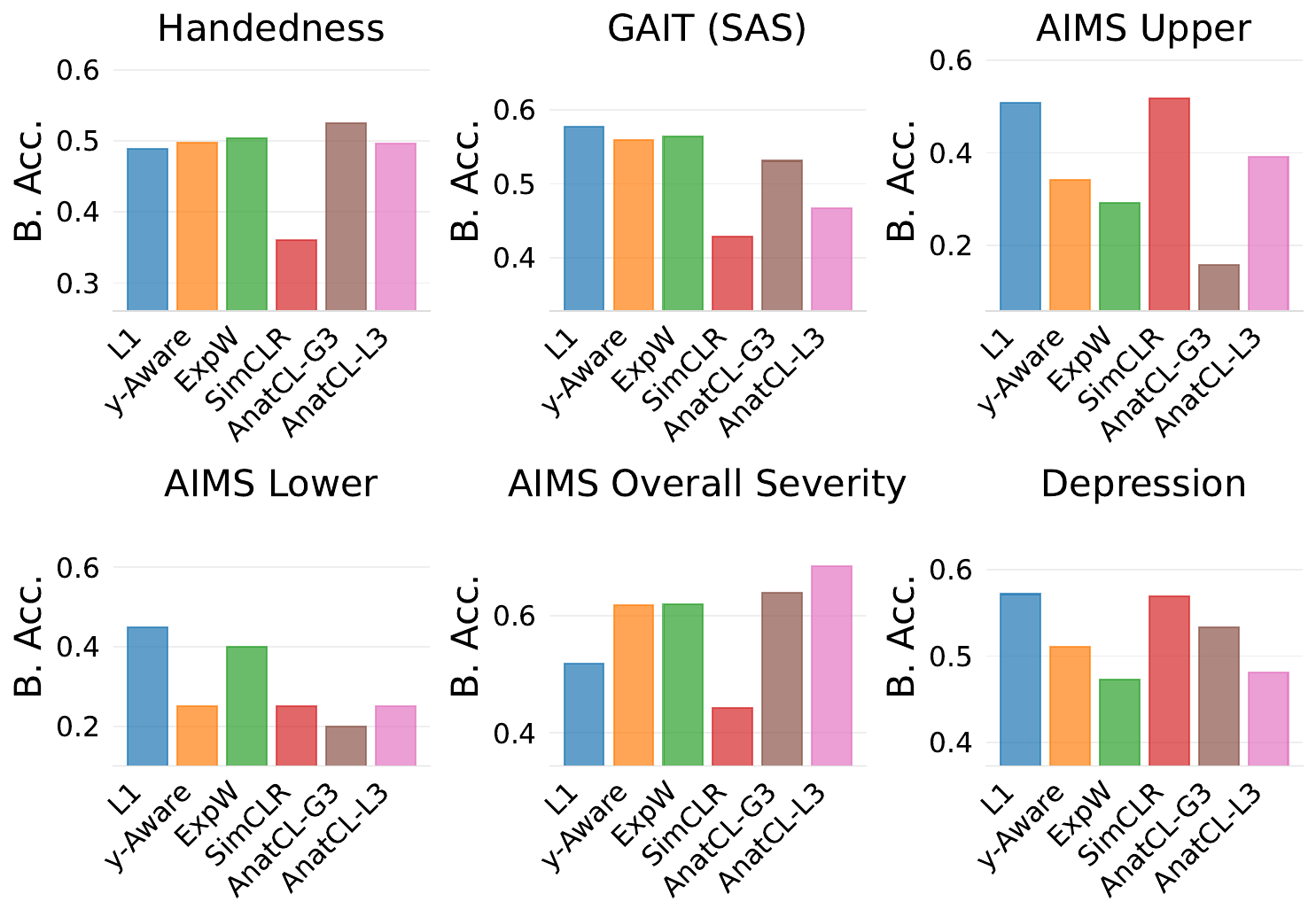}
        \caption{SchizConnect}
        \label{fig:scz-pheno}
    \end{subfigure}
    \caption{\textbf{Prediction of clinical assessment scores} from structural brain MRIs.}
    \label{fig:assessment-results}
\end{figure*}

As final experiments, we turn our attention to predicting clinical assessment scores, cognition, and psychopathology from brain MRIs. To the best of our knowledge, this has not been explored in other works, and it could provide useful insights on the relation between brain anatomy and individual phenotypes, whereas most of the literature focuses on fMRI~\citep{kong2023comparison}. 
The 10 scores considered can be distinguished based on the prediction task: AIMS, depression, handedness, and GAIT are classification tasks, while IQ scores (FIQ, VIQ, and PIQ) are regression tasks. 
For handendess, we predict right-handed vs other (left-handed or ambi), for depression, we classify between absent vs mild and above, for AIMS, we classify between none and minimal vs mild and above, for GAIT, between normal vs everything else. For a more detailed explanation of the considered phenotypes, we refer to the official documentation. %
The results are reported in Fig.~\ref{fig:assessment-results}. 
While we cannot conclude that any of the analysed methods can accurately predict all clinical assessments from structural MRI scans, AnatCL overall achieves the best results three out of ten times, which is more than any other baseline. Interestingly, AnatCL can better predict the overall AIMS score (Fig.~\ref{fig:scz-pheno}), hinting that it may be linked to brain anatomy. This link could lead to the discovery of novel biomarkers and should be studied more in depth in future research.

\section{Conclusions}

We propose AnatCL, a foundation model based on brain anatomy and trained with a weakly contrastive learning approach. With thorough validation on 10 different downstream tasks, we show that incorporating anatomy information during training can result in more accurate predictions of different neurological and psychiatric conditions, and appears promising for clinical assessment scores. We release the weights of the trained model for public use, enabling researchers and practitioners to leverage AnatCL in many different applications.

\noindent\section*{Acknowledgments}
We acknowledge ISCRA for awarding this project access to the LEONARDO supercomputer, owned by the EuroHPC Joint Undertaking, hosted by CINECA (Italy). This work was granted access to the HPC/AI resources of IDRIS under the allocation 2023-AD011013473R1 made by GENCI.
Data collection and sharing for the Alzheimer's Disease Neuroimaging Initiative (ADNI) is funded by the National Institute on Aging (National Institutes of Health Grant U19AG024904). 

\bibliographystyle{model5-names}
\bibliography{refs}

\section*{Supplementary Material}

\section{Limitations, impact, and ethics}
\label{sec:limitations}

We believe that foundation models for neuroimaging may have a considerable impact on accurately diagnosing neurological and psychiatric diseases. With AnatCL we aim at laying the foundations for this path. Currently, AnatCL is limited to using a single data modality (structural MRI), considering limited anatomical features (only CT, GMV, and SA) and to a relatively small backbone (ResNet-18). Future research should focus on improving these issues, in order to obtain even more accurate predictions. Although deep learning techniques can be used in a variety of contexts, we do not believe that AnatCL inherently poses any ethical issue. Furthermore, all the data employed in this work is publicly available to researchers.

\section{Results}

\subsection{Choice of Anatomical Features}

\begin{table*}
    \centering
    \caption{Preliminary study on anatomical features with Ridge regression.}
    \begin{tabular}{l c c c c c c}
    \toprule
         & CT (mean) & GMV & Surf. Area & Int. Mean & Gauss. Curv. Index & Intr. Curv. Index \\
    \midrule
         Neg. MAE & -8.57\std{0.03} &  \textbf{-6.96}\std{0.02} & \textbf{-6.93}\std{0.02} & -7.76\std{0.01} & -8.27\std{0.03} & -8.36\std{0.02} \\
         $R^2$ & 0.32\std{0.00} & \textbf{0.55}\std{0.00} & \textbf{0.55}\std{0.00} &  0.46\std{0.00} &  0.14\std{0.02}  & 0.32\std{0.00} \\
    \bottomrule
    \end{tabular}
    \label{tab:anatomical-features-ablation}
\end{table*}

\begin{table*}[t]
    \centering
    \caption{AnatCL results with all seven anatomical features.}
    \resizebox{\linewidth}{!}{%
    \begin{tabular}{l
                    c c c c | c c c | c c | c}
    \toprule
    & \multicolumn{4}{c|}{\textbf{Schizconnect}}
    & \multicolumn{3}{c|}{\textbf{ABIDE-I}}
    & \multicolumn{2}{c|}{\textbf{ADNI}}
    & \textbf{OASIS-3}\\[-0.2em]
    \textbf{Method}
        & \textbf{SCZ (Broad)}
        & \textbf{SCZ (Strict)}
        & \textbf{Schizoaff.}
        & \textbf{Bipolar}
        & \textbf{Autism}
        & \textbf{Aspergers}
        & \textbf{PDD-NOS}
        & \textbf{HC vs AD}
        & \textbf{sMCI vs pMCI}
        & \textbf{HC vs AD}\\
    \midrule
    SimCLR
        & 58.53\std{3.52} & 68.47\std{8.47} & 51.23\std{11.94} & 67.34\std{9.96}
        & 54.45\std{2.99} & 59.61\std{2.72} & 52.12\std{6.62}
        & 78.47\std{2.51} & 61.77\std{3.85} & 73.97\std{4.98}\\
    L1 (age-sup.)
        & 65.79\std{4.74} & 70.68\std{5.10} & 65.24\std{15.21} & 64.49\std{22.08}
        & 54.53\std{1.79} & \underline{61.33}\std{8.77} & 57.54\std{5.93}
        & \underline{81.20}\std{2.30} & \textbf{68.12}\std{5.42} & 75.40\std{5.40}\\
    y-Aware
        & 66.20\std{4.50} & 71.04\std{2.31} & \underline{70.29}\std{14.73} & 63.95\std{19.81}
        & 55.84\std{3.37} & 60.60\std{9.22} & 56.17\std{10.17}
        & 80.30\std{1.80} & 64.72\std{4.43} & 76.70\std{3.30}\\
    ExpW
        & \textbf{69.53}\std{4.43} & 67.65\std{8.27} & 63.26\std{18.06} & \underline{74.34}\std{18.95}
        & \textbf{58.50}\std{2.41} & 58.68\std{3.82} & {58.31}\std{4.33}
        & \textbf{81.84}\std{2.95} & \underline{66.54}\std{5.64} & 74.67\std{2.87}\\
    \midrule
    AnatCL-L3 (Desikan)
        & 66.03\std{4.06} & 69.86\std{11.20} & 52.72\std{6.64} & 63.12\std{21.50}
        & 53.99\std{2.60} & \textbf{61.44}\std{4.33} & 56.67\std{7.43}
        & 79.71\std{2.51} & 64.78\std{3.57} & 74.26\std{3.32}\\
    AnatCL-L3 (Destrieux)
        & {67.31}\std{5.75} & 70.03\std{7.16} & \textbf{72.47}\std{12.27} & 55.88\std{13.23}
        & 53.14\std{4.20} & 60.99\std{4.94} & 56.63\std{6.74}
        & 79.38\std{3.56} & 64.97\std{3.40} & 77.00\std{2.28}\\
    AnatCL-G3 (Desikan)
        & 64.31\std{4.43} & {72.46}\std{4.41} & 62.42\std{14.95} & 72.42\std{15.77} & \underline{57.12}\std{2.79} & 58.25\std{4.46} & \textbf{64.39}\std{9.26} & 80.84\std{2.38} & 64.88\std{4.37} & \underline{78.32}\std{4.48}\\
    AnatCL-G3 (Destrieux) 
        & 64.12\std{3.18} & \underline{73.04}\std{4.42} & 59.76\std{15.20} & \textbf{84.19}\std{7.10} & 55.41\std{4.55} & 58.96\std{7.52} & 58.10\std{4.63} & 80.75\std{3.35} & 65.91\std{2.23} & \textbf{78.66}\std{2.89}\\
    \midrule
    AnatCL-L7 (Desikan)
        &  66.16\std{6.12} & 70.92\std{8.82} & 63.65\std{18.60} & 66.81\std{20.04} & 55.28\std{1.41} & 60.76\std{7.85} & \underline{59.42}\std{8.02} & 80.17\std{1.98} & 64.59\std{1.75} & 74.22\std{4.21} \\
    AnatCL-L7 (Destrieux) 
        & 68.06\std{4.93} & 70.51\std{10.15} & 63.73\std{17.27} & 60.56\std{14.39} & 55.79\std{2.25} & 59.68\std{1.02} & 56.28\std{3.90} &  81.10\std{2.21} & 63.92\std{5.43} & 77.33\std{4.91} \\
    AnatCL-G7 (Desikan) 
        & \underline{68.16}\std{5.95} & \textbf{74.06}\std{8.34} & 61.97\std{15.20} & 69.33\std{19.83} & 55.60\std{2.37} & 58.04\std{5.92} & 57.15\std{6.08} &  80.79\std{1.32} & 63.46\std{3.46} & 75.69\std{3.22} \\
    AnatCL-G7 (Destrieux) 
        & 66.98\std{3.32} & 72.39\std{6.55} & 59.66\std{16.00} & 67.04\std{14.83} & 54.47\std{3.96} & 60.85\std{5.79} & 55.99\std{6.90} & 79.08\std{2.85} & 64.45\std{2.13} & 75.93\std{3.81} \\
    \bottomrule
    \end{tabular}}
    \label{tab:combined-results-all-features}
\end{table*}

OpenBHB provides 7 different anatomical measures, computed with FreeSurfer: Cortical Thickness (CT, mean and std values), Gray Matter Volume (GMV), Surface Area, Integrated Mean, Gaussian Curvature Index, and Intrinsic Curvature Index. %
The features we select are inspired by the MIND metric~\citep{sebenius2023robust}.
To further motivate the choice of the anatomical features, we conduct a preliminary study by training a simple linear model (Ridge regression) over each feature, with the task of brain age prediction on the OpenBHB data. The results are evaluated in terms of negative mean absolute error and $R^2$ score and are reported in Tab.~\ref{tab:anatomical-features-ablation}. The best results are achieved by GVM and Surface Area, which are morphological features regarding the volume and the total area of each ROI. Thus, we select these two features for our experiments with AnatCL.
In addition to GMV and Surface Area measures, we also include CT as it is a widely used measure in related works~\citep{sihag2024explainable, sihag2024towards}. The final choice is thus composed of three features, which cover different aspects of the morphology and geometry of brain ROIs (e.g., cortical thickness is relevant for conditions such as Alzheimer's Disease~\citep{phan2024increased}, and volume and surface measurements can signal brain atrophy, linked to different conditions). 

In addition, we also report results of AnatCL employing all seven available measures in Tab.~\ref{tab:combined-results-all-features}. As we can observe, using just the chosen features or all available ones achieves comparable results, with some strongest peak in the reduced version.

\subsection{Ablation study}

In this section, we perform an ablation study to motivate the design choices we made for AnatCL. Our ablation study aims to assess the benefit of different network architectures (i.e. larger models), and different formulations of the AnatCL loss function.

\paragraph{Model size and architecture}
We perform an ablation on different neural architectures, considering larger models such as 3D ResNet-50. We report the results of this ablation on the Schizconnect dataset here in Tab.~\ref{tab:ablation-study-resnet}.
From the obtained results, we do not find significant advantages in using a 3D ResNet-50 over the simpler and smaller version 3D ResNet-18. For this reason, we chose to employ this simpler architecture, as it is practically more convenient to train and deploy in real settings. 
It is worth mentioning that similar findings are reported in related literature~\citep{dufumier2022openbhb}, showing that deeper architectures such as even a 3D DenseNet-121 do not bring significant improvements over a 3D ResNet-18, at least on VBM-processed images. We did not explore different and larger architectures such as ViT~\citep{dosovitskiy2020image}, as we were limited by computational constraints. In fact, a 3D ViT implementation requires significantly more memory and computational power, which makes it less practical and favorable with respect to ResNet-18, also from a carbon footprint point of view.

\begin{table*}
    \centering
    \caption{Model size ablation study. Larger architectures do not result in consistent improvements over simpler models, as also shown in~\cite{dufumier2022openbhb}.}
    \begin{tabular}{l l c c c c}
    \toprule
    \textbf{Model} & \textbf{Method} & \textbf{SCZ. (Broad)} & \textbf{SCZ. (Strict)} & \textbf{Schizoaff.} & \textbf{Bipolar}  \\
    \midrule
    ResNet-18 & y-Aware & 66.20\std{4.50} & 71.04\std{2.31} & \underline{70.29}\std{14.73} & 63.95\std{19.81} \\
    & ExpW & \textbf{69.53}\std{4.43} & 67.65\std{8.27} & 63.26\std{18.06} & {74.34}\std{18.95} \\
    & AnatCL-L3 & 66.03\std{4.06} & 70.03\std{7.16} & \textbf{72.47}\std{12.27} & 63.18\std{21.50}\\
    & AnatCL-G3 & 64.31\std{4.43} & \underline{73.04}\std{4.42} & 62.42\std{14.95} & \textbf{84.19}\std{7.10} \\
    \midrule
    ResNet-50  & y-Aware & 66.14\std{9.23} & 68.17\std{5.45} & 64.94\std{17.13} & 69.16\std{22.00}\\
    & ExpW & 66.39\std{6.54} & \textbf{73.69}\std{5.23} & 66.07\std{14.89} & 66.19\std{19.95} \\
    & AnatCL-G3 (Desikan)& \underline{68.07}\std{4.16} & {72.16}\std{5.53} & 57.92\std{14.50} & \underline{78.71}\std{18.67} \\
    & AnatCL-L3 (Desikan) & 65.20\std{3.19} & 62.99\std{8.39} & 69.40\std{9.98} & 67.61\std{15.81} \\
    \bottomrule
    \end{tabular}
    \label{tab:ablation-study-resnet}
\end{table*}

\paragraph{Loss formulation and pretraining}
We also experiment with different variants of the AnatCL loss. First, we set $\lambda_2 = 0$, meaning that we only consider anatomical features in the training process. We call this variant AnatSSL (as in Self-Supervised), as it effectively does not require any other information besides the anatomical data itself. This variant may be useful when pretraining on a large cohort in which the patient's age is unknown. We also perform a simple pre-training of the network using an L1 loss to predict the anatomical features: 
\begin{equation}
    \mathcal{L}_{anat-sup} = || r(f(x_i)) - \bar{\omega}_i||
\end{equation}
\noindent where $r: \mathbb{R}^d \rightarrow \mathbb{R}^3$ is a linear regression layer, and $\bar{\omega}_i$ is the average of the patient's global descriptors: 
\begin{equation}
    \bar{\omega}_i = \left[\frac{1}{68}\sum_j \omega^1_{i,j}, \dots,  \frac{1}{68}\sum_j \omega^3_{i,j} \right]
\end{equation}

We consider the average global descriptor composed of 3 values for this formulation, instead of the average local vector which would contain 68 elements for regression for practical reasons. The results are presented in Tab.~\ref{tab:loss-ablation}.
\begin{table}
    \centering
    \caption{Loss function and pre-training strategy ablation. AnatCL learns more generalizable representations by combining anatomical features and patients' metadata (i.e. age).}
    \resizebox{\linewidth}{!}{%
    \begin{tabular}{l c c c}
     \toprule
     & \multicolumn{2}{c}{\textbf{ADNI}} & \textbf{OASIS-3} \\
     \textbf{Method} & \textbf{HC vs AD} & \textbf{sMCI vs pMCI} & \textbf{HC vs AD} \\
     \midrule
     L1 (anat. supervised) & 64.63 & 60.79 & 67.01 \\
     AnatSSL-L3 & 76.32 & 61.11 & 69.12 \\
     AnatSSL-G3 & 73.21 & 60.69 & 69.77 \\
     AnatCL-L3 & \underline{79.71} &\underline{64.78} & \underline{74.26} \\
     AnatCL-G3 & \textbf{80.84} & \textbf{64.88} & \textbf{78.32} \\
     \bottomrule
    \end{tabular}
    }
    \label{tab:loss-ablation}
\end{table}
As we observe from the results, employing only the anatomical features is not enough to obtain more robust representations. However we can still notice how contrastive-based methods such as AnatSSL outperforms traditional supervised baselines.
Our proposed approach, on the other hand, leverages information from both the patient and brain anatomy, leading to more robust and generalizable representations.

\subsection{Assessment scores and phenotypes}

Full numerical values are presented in Tab.~\ref{tab:assessment-results}.

\begin{table*}
    \centering
    \caption{Assessment scores / phenotypes prediction from brain MRIs.}
    \resizebox{\linewidth}{!}{%
    \begin{tabular}{l c c c c c c | c c c c}
    \toprule
    & \multicolumn{6}{c|}{\textbf{SchizConnect}} & \multicolumn{4}{c}{\textbf{ABIDE}} \\
     \textbf{Method} & \textbf{AIMS Overall} & \textbf{AIMS Up.} & \textbf{AIMS Low.} & \textbf{Depression} & \textbf{Handedness} & \textbf{GAIT} & \textbf{Handedness} & \textbf{FIQ (MAE)} & \textbf{VIQ (MAE)} & \textbf{PIQ (MAE)} \\
    \midrule
    SimCLR & 44.33\std{29.92} & \textbf{51.67}\std{16.16} & 25.00\std{27.39} & \underline{56.93}\std{13.88} & 36.06\std{2.72} & 42.83\std{9.61} & 
    49.26\std{5.78} & 84.65\std{16.36} & 89.07\std{15.14} & 89.68\std{15.00} \\
    L1 & 51.83\std{24.43} & \underline{50.83}\std{20.82} & \textbf{45.00}\std{33.17} & \textbf{57.17}\std{13.06} & 48.91\std{5.60} & \textbf{57.73}\std{8.64} & 48.18\std{9.43} & 42.16\std{31.17} & 46.54\std{32.57} & 46.64\std{32.24} \\
    y-Aware & 61.83\std{15.87} & 34.17\std{11.90} & 25.00\std{27.39} & 51.09\std{5.32} & 49.71\std{7.69} & 55.92\std{9.52} & \underline{49.45}\std{1.56} & \textbf{42.10}\std{31.14} & 45.38\std{33.19} & 45.35\std{32.76} \\
    ExpW & 62.00\std{12.40} & 29.17\std{17.08} & \underline{40.00}\std{33.91} & 47.26\std{7.27} & \underline{50.39}\std{6.28} & \underline{56.39}\std{14.14} & 49.53\std{3.26} & 42.94\std{30.76} & \textbf{45.28}\std{33.23} & \textbf{45.02}\std{32.88} \\
    \midrule
    AnatCL-G3 & \underline{64.00}\std{12.72} & 15.83\std{12.19} & 20.00\std{18.71} & 53.35\std{8.54} & \textbf{52.44}\std{9.14} & 53.10\std{11.69} & \textbf{50.21}\std{6.82} & 45.18\std{29.68} & 49.07\std{31.52} & 49.30\std{30.86}  \\
    AnatCL-L3 & \textbf{68.50}\std{18.09} & 39.17\std{14.81} & 25.00\std{27.39} & 48.05\std{10.89} & 49.67\std{8.06} & 46.74\std{5.05} & 49.13\std{3.32} & 50.77\std{27.14} & 56.18\std{28.06} & 56.13\std{27.62} \\
    \bottomrule
    \end{tabular}
    }
    \label{tab:assessment-results}
\end{table*}

\end{document}